# A Case Study on the Performance Metrics of Integrated Photonic Computing

Frank Brückerhoff-Plückelmann[1,2], Jelle Dijkstra[2], Julian Büchel[1], Bottyan Batkai[2], Falk Ebert[2], Luis Mickeler[3], Urs Egger[2], Abu Sebastian[1], Wolfram Pernice[2], Ghazi Sarwat Syed[1]

[1]IBM Research – Europe, 8803 Rüschlikon, Switzerland
[2]University of Heidelberg, Kirchhoff-Institut für Physik, 69120 Heidelberg, Germany
[3]Physical Institute, University of Münster, Heisenbergstraße 11, 48149 Münster, Germany

**Abstract**

Photonic processors use optical signals for computation, leveraging the high bandwidth and low loss of optical links. While many approaches have been proposed, including in-memory photonic circuits, most efforts have focused on the physical design of photonic components rather than full architectural integration with electronic peripheral circuitry. In this work, we present a microarchitecture-level study that estimates the compute efficiency and density of three prominent photonic computing architectures. Our evaluation accounts for both the photonic circuits and the essential peripheral electronics required for opto-electronic and analog–digital conversions. We further demonstrate a heterogenous photonic–electronic in-memory processing approach for low-latency neural network inference. These results provide a better understanding of the design aspects of photonic computing.

**Introduction**

The demand for computational power is growing at an unprecedented rate, especially driven by the increasing complexity of artificial intelligence (AI) applications. In general, these models comprise a composition of linear and non-linear functions, enabling them to serve as highly expressive systems capable of approximating complex relationships within data. Emerging hardware architectures increasingly aim to co-locate memory and linear processing, or at least bring them sufficiently closer together, to circumvent the memory bottleneck. To meet these demands, specialized digital accelerators have been developed, such as systolic arrays for matrix multiplications[1]. These accelerators have been instrumental in enabling state-of-the-art AI models, with profound implications for various fields, including scientific breakthroughs such as protein structure prediction with AlphaFold[2]. However, despite their success, these digital accelerators are highly power-intensive, and the latest AI models continue to push them to their limits.

Inspired by the remarkable efficiency of analog computing in biological brains, exploring analog processors as the next generation of hardware accelerators is intriguing. These accelerators do not rely on abstract digital encodings but directly harness the physical behavior of devices and circuits. Data is directly encoded in physical quantities, e.g. the power of on optical pulse, and computation is implemented by manipulating these quantities[3], e.g. with tunable absorbers. In the electronic domain, such accelerators have already reached a level where they can perform real-world tasks at the system level[4]. Meanwhile, photonic analog processors, though still in an earlier stage of architectural and integration development[5–8], offer unique advantages over their electronic counterparts such as higher bandwidths and lower transmission losses. There is a wide range of envisioned architectures and applications for photonic processors. Free-space optical computing naturally enables large-scale matrix operations due to its three-dimensional nature but faces challenges in fabrication scalability compared to integrated solutions[9,10]. Input/output interfaces also vary significantly between approaches. For instance, directly passing signals from a sensor into a photonic processor could greatly enhance performance by enabling analog-domain processing of conventionally digital signal processing tasks[11,12].

However, performance estimates often assume ideal, seamless electronic-digital interfaces, and large-scale integration. In practice, in addition to analog-to-digital conversions, input/output operations also require optoelectronic conversions, which impact overall compute performance. Considering just that, in this work, we provide an optimistic but realistic performance estimate of photonic computing schemes. We focus on three prominent linear photonic processor architectures, namely microring (MRR) weight banks, photonic crossbars and Mach-Zehnder Interferometer (MZI) meshes which have indicated potential for full on-chip integration and are also pursued commercially. We begin by examining the key components contributing to the peripheral overhead of photonic accelerators, with a particular focus on the power and area costs associated with analog-to-digital and digital-to-analog conversions. Next, we compare various mapping techniques for encoding signed weights and inputs. We then assess the computational performance of these photonic architectures. Along the way, we identify several fundamental challenges rooted in the physics of optics that limit photonic processor design. Finally, we explore opportunities to harness the benefits of photonic computing while mitigating its constraints and demonstrate how hybrid systems combining photonic and electronic in-memory processors can be realized.

**Results**

Interfacing analog photonic processors with today's predominant digital and electronic infrastructure presents several challenges. Optical carrier signals must be generated, and digital electronic signals must be converted and coupled to the optical wave. Similarly, after photonic processing, the optical signal must be converted back to an electronic one and mapped to the digital encoding scheme, as illustrated in Figure 1.

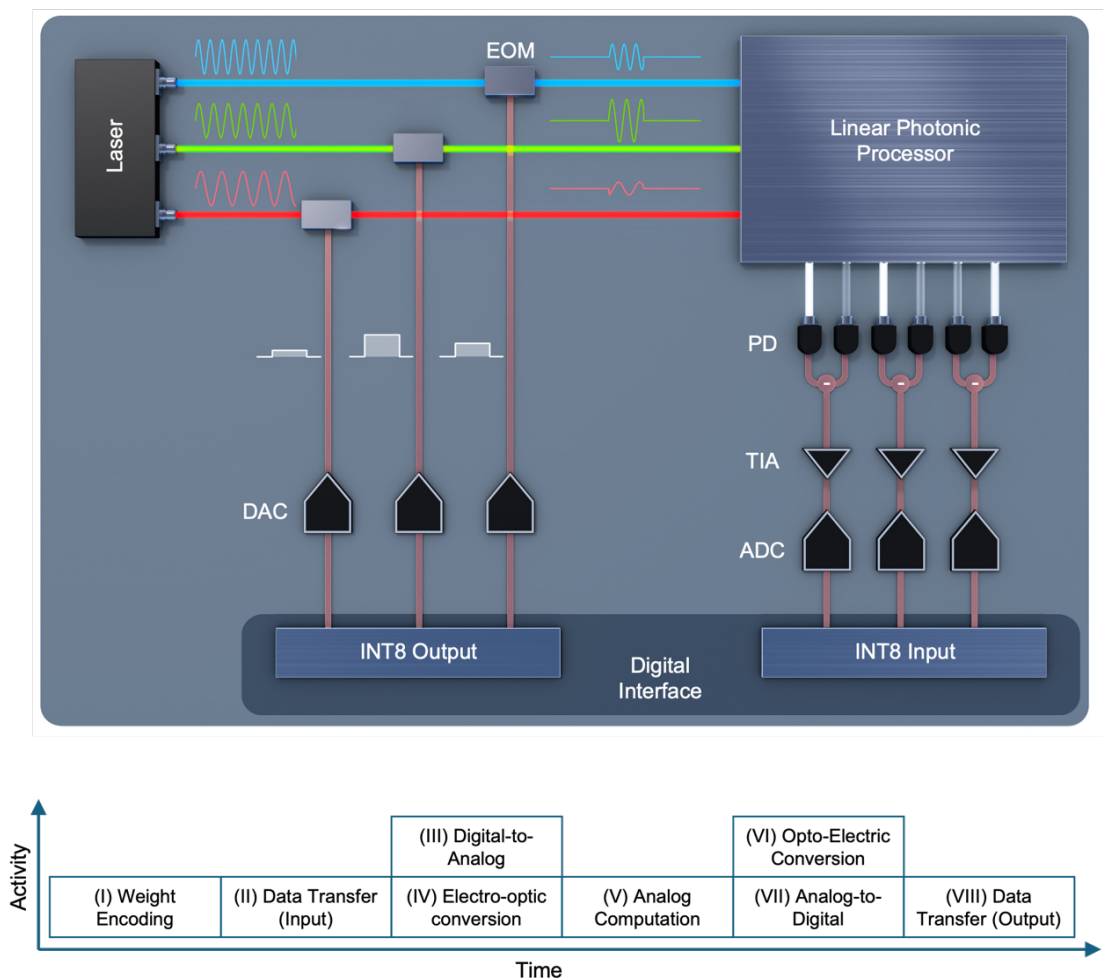

*Figure 1. System architecture of integrated linear processors.* The digital interface outputs signed INT8 values, which are converted into electronic pulses via pulse amplitude modulation in the digital-to-analog converters (DACs). Electronic optic modulators (EOMs) couple these pulses to high-frequency optical carrier signals. The photonic circuit processes the encoded inputs, and photodetectors (PDs) convert the optical signals back

*into the electronic domain. A balanced readout scheme enables negative weights. Transimpedance amplifiers (TIAs) convert and amplify the PD output currents into the voltage domain. Finally, analog-to-digital converters (ADCs) map the analog signals back to signed INT8 values before forwarding them to the digital interface.*

To make different processor architectures comparable, we assume pulse amplitude modulation, encoding around the electro-optic modulator's bias point and incoherent detection. At the output, we assume a balanced detection scheme to enable negative weights and require the photonic processor to provide a signal swing 50x larger the input referred noise of the readout electronics, enabling accuracy comparable to a digital system with 4bit weight quantization. The output electronics also feature a high-pass filter, to remove both low-frequency noise and to enable negative input values[7,13]. We assume ideal passive components in the photonic routing and only consider the architectural choice and the impact of the memory units. The individual simulation steps are explained in detail in the Methods section.

Microring Weight Banks

Microring resonator (MRR) weight banks deploy a broadcast and weight architecture as sketched in Figure 2a. The inputs, encoded on different wavelengths, are first multiplexed together and then evenly distributed to N weight banks. Within each weight bank, tunable add-drop ring resonators, one per wavelength channel, store the weight information. Individually tuning the resonance wavelength of the ring resonators enables an arbitrary splitting between the positive and negative output for each vector component[6,14].

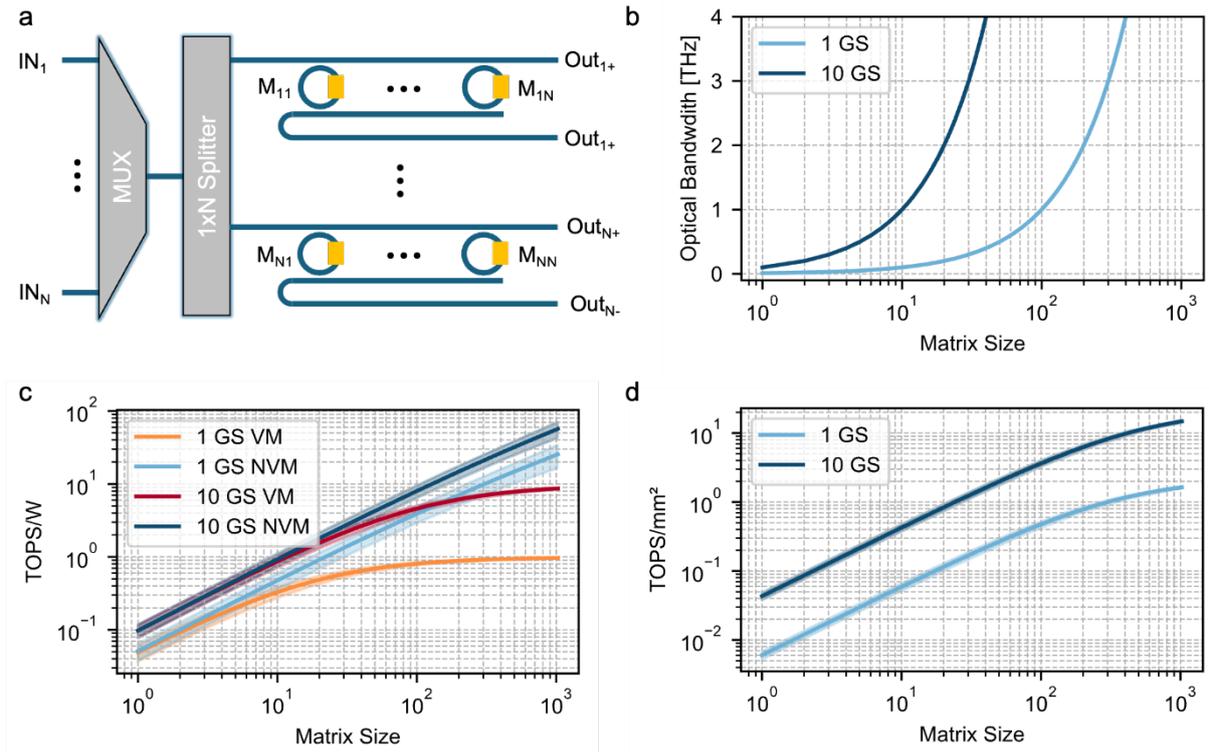

*Figure 2. Performance of microring resonator banks. a, The input signals are combined with a multiplexer (MUX) and simultaneously broadcasted to all weight banks. Within each bank, an add-drop filter is assigned to each input wavelength, splitting the signal between the positive and negative outputs. b, The total available optical bandwidth limits the scalability of MRR architectures. To obtain linear weighting, each resonator's bandwidth must significantly exceed the signal bandwidth. Typical free spectral ranges of silicon add-drop filters are around 1 terahertz and the full optical C-band spans 4 terahertz, placing an upper bound on the matrix size. c, For non-volatile memory (NVM) computational efficiency steadily increases with matrix size whereas the static power consumption of volatile memory (VM) limits the achievable efficiency. d, Computational density saturates at a value determined by the physical size of a single weight, which is similar for VM and NVM*

> *approaches. As the photonic circuit becomes the dominant contributor to area, increasing the operating frequency enhances computational density.*

The available optical bandwidth and free spectral range (FSR) of the ring resonators constrain the system's scalability. Modulating the optical carrier with a sampling frequency of $f_s$ increases the bandwidth of the carrier by $f_s$, as mixing the carrier with the electronic signal creates a sideband on each side with a bandwidth given by the electronic bandwidth of $f_s/2$. The full width half maximum of the ring resonances must be substantially larger, i.e. around M=10 times the optical bandwidth[15], to obtain linear weighting functionality without distorting the pulse shape. Thus, for an NxN MRR weight bank, the FSR of the ring resonators must be at least:

$$\text{FSR} \geq N \cdot M \cdot f_s$$

The radius of the ring resonator determines the FSR and thus the number of wavelength channels that can be deployed without crosstalk. While bend radii as low as 3 µm are possible on the silicon on insulator (SOI) platform[16], practical implementations typically deploy larger resonators as phase tuning benefits from longer propagation length. Figure 2b shows the required bandwidth for operation at 1 GS/s and 10 GS/s per second. Due to the limited FSR of MRRs and eventually the limited overall optical bandwidth, fast operation speeds are not compatible with matrix sizes in the order of N = 100 or larger. For example, E. Blow et al. realized a four-resonator weight bank using 11.3 µm rings with a FWHM of 43.37 GHz for high-speed operation resulting in an FSR of 1.07 THz, limiting the theoretical scalability to 24 channels[15]. For all photonic architectures, there are three main contributions to the total power consumption. The power consumption of the electronic interface increases linearly with matrix size, the static power consumption of the volatile memory (VM) scaling quadratically with matrix size and the total laser pump power, which scales like $O(N^{1.5})$ for MRR banks. Figure 2c illustrates the computational efficiency for both a non-volatile phase change material (PCM) based phase shifter[17] and a heater with 2 mW power consumption[18]. Since low-loss PCMs are a less mature technology, we assume a limited resonance tuning capability, i.e. a transmission on -1 dB in the high state and -11 dB in the low state[17], and thus a limited memory window. MRR weight banks have the potential to reach computational efficiency beyond 10 TOPS/W, but this requires matrix sizes exceeding N = 100 and matrix weights without static power consumption. Since both weighting schemes eventually require a ring resonator and electrodes for PCM switching or applying current to the heater, we assume 900 µm$^2$ per matrix element for both for VM and NVM. Figure 2d shows the computational density in dependence on the matrix size. The computational density is limited by the electronic interface for small matrix sizes and by the single weight area for large sizes.

Crossbar Arrays

Photonic crossbar arrays distribute each input pulse evenly across all output waveguides, with the transmission along each path individually programmable via tunable absorbers to encode matrix elements. For balanced readout, two schemes can be employed: a single reference output[19] or a pairwise output configuration[20]. While the single reference scheme offers a more compact design and minimizes the fan-out loss by halving the circuit size, it only offers half the memory window. We consider this design for further analysis, as illustrated in Figure 3a.

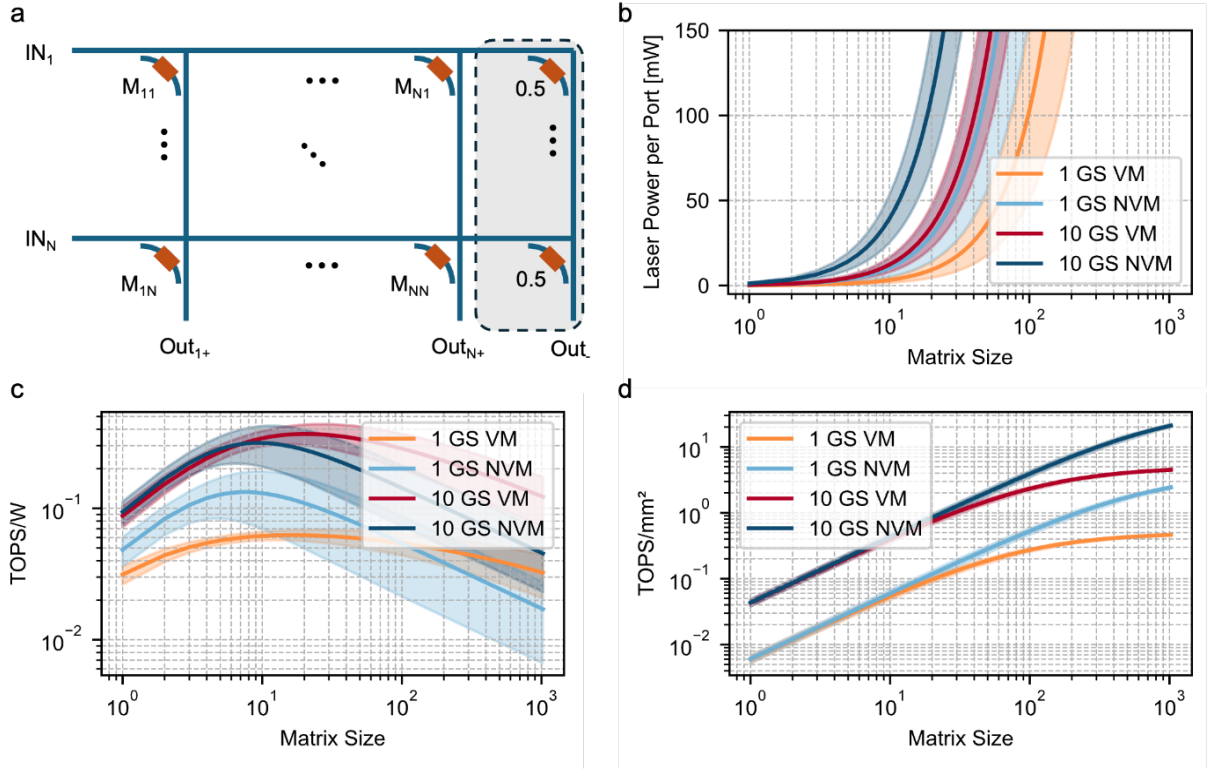

*Figure 3. Performance of photonic crossbar arrays.* **a,** Photonic crossbars use a balanced passive transmission matrix based on a beam-splitter network. Tunable absorbers in each matrix cell enable independent programming of the matrix weights. A single computed reference output serves as a common negative reference. **b,** For an N x N matrix, the laser power per carrier increases with $N^{1.5}$. Given typical on-chip laser power limits in the order of 100 milliwatts, this scaling constrains the achievable matrix size. **c,** High optical power demands lead to low computational efficiency compared to other architectures. The efficiency approaches zero for large matrices due to the more-than-quadratic scaling of total input power, independent from the memory type. **d,** Due to the compact size of absorption-based NVM, photonic crossbars can potentially achieve high computational densities, especially for fast operation as the photonic circuit size is independent from the processing speed.

The photonic crossbar performs accumulation using passive broadband couplers, which makes the system inherently lossy. For an NxN crossbar, the matrix size dependent transmission $\alpha_{XBar}$ from one input to one output is:

$$\alpha_{XBar} = \frac{1}{N^2}$$

Despite this loss, the architecture benefits from wavelength independence, ensuring excellent stability and compatibility with wavelength-division multiplexing schemes. Moreover, the option to use the same optical carrier for all inputs eliminates bandwidth constraints related to matrix size[7]. However, the lossy nature of the architecture eventually limits the scalability due to limited on-chip laser power. Figure 3b shows the required laser power per input for non-volatile memory based on the PCM Germanium-Antimony-Tellurium with a memory window of 0.26[21] and a volatile optical attenuator on silicon with a power consumption of 25 mW and a memory window of 0.83[22,23]. The low memory window of the NVM is mostly due to large insertion losses in the high transmission state, e.g. caused by propagation loss in the amorphous state and limited switching volumes. Due to high power demand, larger matrix sizes are only feasible for smaller operation speeds that require a smaller optical output power swing. For example, Bowei et al. demonstrated a 9x3 crossbar array operating with an interface speed of 2 GS/s[7]. As for the MRR banks, there are the interface, weight and laser power contributions to the total power consumption. However, due to the lossy architecture the total pump power increases like $O(N^{2.5})$. This strongly limits the peak computational efficiency and especially pushes it to zero for large matrix sizes, as shown in Figure 3c. As absorptive NVM can be directly placed on top of the waveguide without additional Mach-Zehnder interferometer like structures, very compact memory cells are possible[24]. In addition, multiplexing techniques, that increase the interface size but not the photonic crossbar area can further enhance computational density[20]. Figure 3d shows the computational density, assuming 500 µm$^2$ per NVM matrix cell and 4000 µm$^2$ per VM matrix cell [23].

Mach-Zehnder Interferometer Meshes

Mach-Zehnder Interferometer (MZI) meshes implement the matrix weights using a series of tunable 2×2 optical splitters[5]. There are two modes of operation. In the first, a single coherent optical carrier is used across all inputs, enabling manipulation of both phase and amplitude unlike intensity-only encoding schemes[25]. However, processing and detecting complex-valued signals requires coherent detection schemes, which increases circuit complexity and deviates from the functionality of standard hardware accelerators. Instead, we focus on an incoherent MZI-mesh architecture[26], illustrated in Figure 4a, which assigns a different carrier wavelength to each input signal. Each 2x2 splitter contains up to two phase-shifter, one in each arm, and thus has a programmable transfer function. By placing at maximum N+1 tunable splitters in a row (optical depth), using in total N$^2$ phase shifters and deploying one reference row similar to photonic crossbars, arbitrary real-valued MVM can be performed [26].

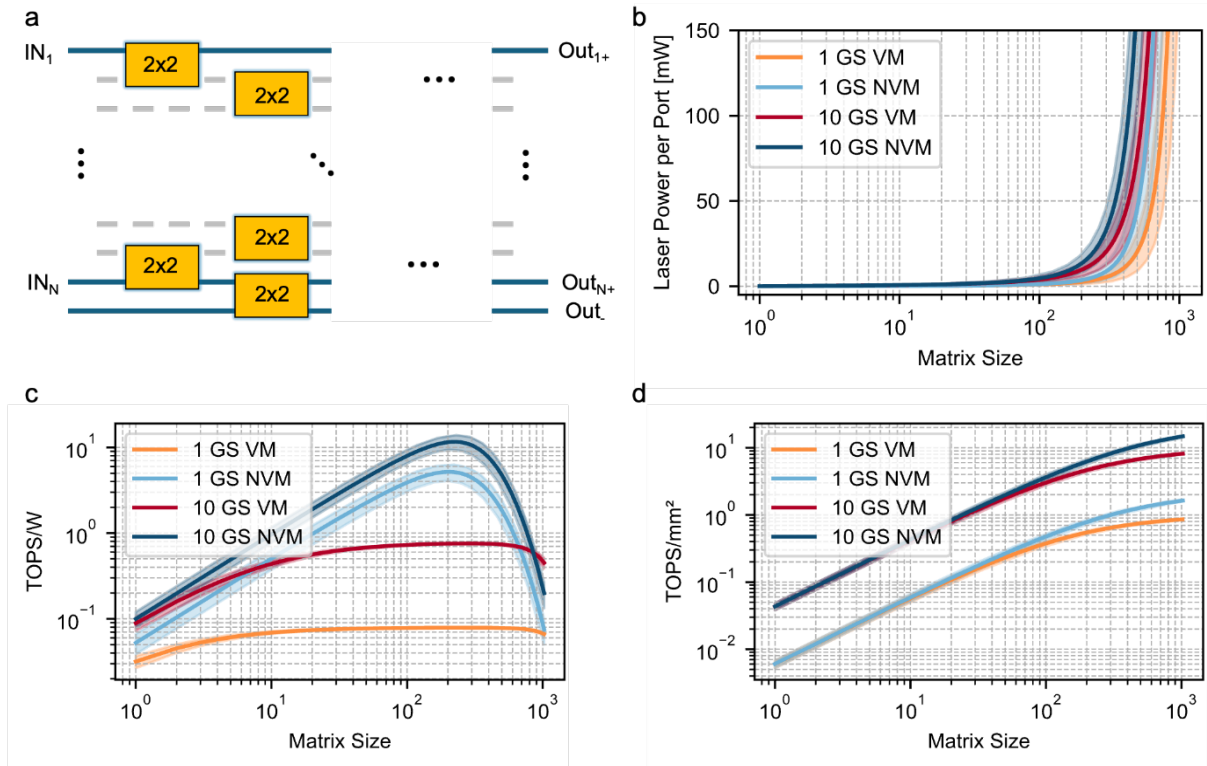

*Figure 4. Performance of Mach-Zehnder Interferometer (MZI) meshes. a, MZI meshes perform optical matrix multiplication using cascaded tunable 2×2 splitters. Only incoherent power-encoded schemes are considered here. b, Scalability is primarily limited by the optical depth, as each signal traverses multiple tunable splitters, accumulating loss. With ultra-low-loss components, matrix sizes exceeding 100×100 are achievable. c, Utilizing non-volatile phase shifters enhances computational efficiency at larger scales. However, efficiency eventually drops to zero due to exponential increase of insertion loss. d, As for the other architectures, the computational density is limited by the interface area for smaller matrix sizes and by the single weight area for large matrix sizes.*

The primary limitation of this architecture lies in the exponential increasing optical loss, as signals must pass through multiple programmable splitters. For an NxN MZI mesh with a splitter transmission of $α_{Sp}$, the matrix size dependent transmission is:

$$α_{MZI} = \frac{1}{N} \cdot α_{Sp}^{N+1}$$

Figure 4b presents the per-input laser power requirements for two tuning mechanisms: heater-based and PCM-based phase shifters. Although the difference in insertion loss is small, 0.23 dB for active tuning [27] and 0.3 dB for non-volatile tuning[17], it has a substantial impact on total power requirements. Matrix sizes above 100x100 are feasible, if the 2x2 splitters are optimized for low insertion loss. Lightmatter for example presented a 64x64 programmable MZI-Mesh for computing[28]. However, a practical constraint is potential phase errors due to fabrication errors requiring additional correction schemes [29]. Figure 4c shows the computational efficiency assuming no static power consumption for the non-volatile PCM phase shifters and 25 mW per heater-based ones[27]. In contrast to the other photonic architectures, the loss increases exponentially with matrix depth. The exponential increase eventually limits the scalability and lets the computational efficiency converge to zero for very large matrix sizes. To estimate the computational density, we assume 900 µm$^2$ per non-volatile phase-shifter and 2000 µm$^2$ for a heater based one[17,27]. Since the chip size is dominated by the photonic matrix weight for large matrixes, the computational density plateaus.

**Discussion**

Hybrid electronic-photonic architectures, such as Lightmatter's quad-core 128×128 matrix processor[8], and purely electronic systems like IBM's 64-core 256×256 design[4] achieve state-of-the art performance for analog computing and demonstrate full integration. Pure photonic computing remains limited in scalability, mostly due to the lack of integration. Table 1 shows a comparison between industrial hardware accelerators and the simulated characteristics of photonic processors. While limited, there exist approaches to increase the sizes of MRR weight banks and crossbar arrays. For ring resonator-based systems, increasing the available bandwidth is crucial. Promising approaches include eliminating the free spectral range (FSR) constraint by combining ring resonators with Bragg gratings[30]. In photonic crossbars, reducing laser power consumption is essential, necessitating a less lossy signal superposition. One potential solution is using ring resonators for accumulation following tunable attenuation, as shown by Varri et al[31]. Since the rings in this configuration are used solely for transmission rather than weighting, the required resonance width is relaxed. MZI meshes, while showing the best scalability in this analysis, are primarily limited by fabrication imperfections. Due to the high optical depth, small deviations in the 2×2 tunable splitters can lead to significant performance degradation. Although compensation schemes exist, they inherently reduce system efficiency[29]. Post-fabrication tuning and broader industrial adoption of integrated photonics could improve yield and consistency[31].

|  | Size | Efficiency | Density | Latency |
|---|---|---|---|---|
| **IBM Hermes**[4] | 256x265 | 9.76 TOPS/W | 1.55 TOPS/mm$^2$ | 127 ns |
| **Lightmatter**[8] | 128x128 | 0.81 TOPS/W | 0.047 TOPS/mm$^2$ | 1-10 ns* |
| **MRR weight banks**[6,15] | ≤ 100x100 | ≤ 15 TOPS/W | ≤ 0.5 TOPS/mm$^2$ | potentially below 1ns |
| **Crossbar arrays**[19,32] | ≤ 100x100 | ≤ 0.4 TOPS/W | ≤ 2 TOPS/mm$^2$ | |
| **MZI-Meshes**[5,28] | ≤ 800x800† | ≤ 10 TOPS/W | ≤ 10 TOPS/mm$^2$ | |

*Table 1. Performance comparison of representative computing platforms. IBM's Hermes chip employs analog in-memory computing using non-volatile phase-change materials, while Lightmatter's platform utilizes optical broadcast with electronic weighting and accumulation. The values shown for the three photonic platforms are extrapolated based on specific assumptions and architectural configurations; alternative design choices may yield different results. *The latency depends on the exact ADC architecture. †Eventually limited by chip size and fabrication imperfections.*

In terms of power efficiency, MRR weight banks and, to a lesser extent, MZI meshes exhibit favorable scaling, potentially exceeding 10 TOPS/W. In contrast, photonic crossbars currently lack competitive power scaling. It is important to note that analog systems are benchmarked under conditions mimicking digital behavior, such as INT8 input/output quantization and an equivalent weight quantization of approximately 4 bits. Adjusting weight precision directly affects computational efficiency across both analog and digital domains. Computational density across all architectures is comparable to other hardware accelerators when either the interface supports high sampling rates or when lower sampling rates are paired with large matrix dimensions and small memory sizes such as in photonic crossbars with non-volatile memory NVM devices. Furthermore, photonic crossbars support WDM, which could enhance density further[20]. Regarding memory types, NVM, due to its zero static power consumption, offers significant advantages over VM-based approaches such as thermal tuning, by reducing overall power draw. It may also simplify input/output design for weight programming. The development of compact electro-optic VMs without static power consumption, such as those based on BTO, could substantially improve the computational efficiency of VM-based photonic accelerators[33]. Finally, while this study focuses on metrics such as computational efficiency and density, most relevant to parallel, high-throughput computing, it is crucial to highlight the inherent advantage of photonic computers in executing full matrix-vector multiplications at high bandwidth. This enables ultra-low latency, which is particularly valuable for iterative computing tasks[34]. Photonic computing also uniquely supports integration of entropy sources for probabilistic computing[35,36], can solve optimization[37] and correlation detection[38] problems, and implement associative memories[39].

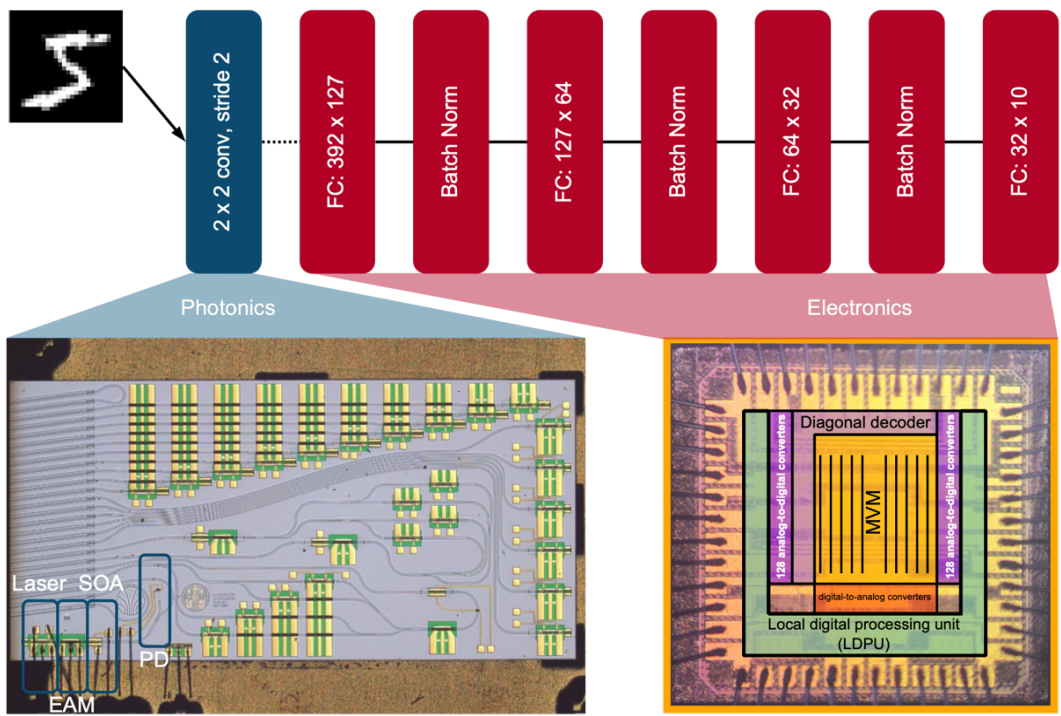

*Figure 5. Heterogenous Neural Network Inference. While photonic computing can enable energy efficient ultra-low latency computing, providing the required memory capacity and high throughput is challenging. Thus, a heterogenous computing framework becomes especially compelling. As a demonstrator, we illustrate this with a photonic–electronic in-memory processing approach. We use a photonic multiplier on Indium Phosphide featuring an electro absorption modulator (EAM) for input encoding and a semiconductor optical amplifier (SOA) for weighting in combination with a phase change material based electronic in-memory computing. After hardware aware training with the AIHWKIT-lightning package[40], the analog system exhibits a classification accuracy of 97.7%, approximately one percent lower than the digital performance.*

Overall, high-throughput applications remain a significant challenge for photonics, its strengths, such as high bandwidth, low latency, and unique features like WDM, make it highly promising for specific problem domains. Thus, only offloading certain computation heavy tasks to photonics while keeping the rest of the system in the electronic domain, as sketched in Figure 5, is especially promising. One prominent example is neural network inference. In convolutional neural networks (CNNs), for an image of size $n \times n$ and a filter of size $k \times k$, the number of multiply–accumulate operations required scales as $(n - k)^2$, where $n \gg k$ is a typical case for the first layer. Since $k$ is small, the first layer can be efficiently mapped onto scalable photonic circuits. However, subsequent layers, including dense layers that require lots of computational memory, are not easily accommodated in the photonic domain. We demonstrate this using a combination of mixed-signal photonic in-memory computing based on VM devices for the first layer, and NVM-based electronic in-memory computing implemented on the IBM HERMES Project chip[41] for deeper layers. Further work is, however, needed to unlock the full potential of this approach.

In summary, using a custom performance benchmarking framework, we have evaluated the compute efficiency and density scaling of several prominent photonic computing architectures and compared them with electronic processors. We have outlined the fundamental advantages and limitations of each architecture and demonstrated a heterogeneous computing framework to guide the optimal use of photonic computing.

**Methods**

Output Distribution Statistics and Quantization

The ratio between signal and noise at the output of the photonic processor will eventually determine the overall accuracy. Considering an optical input power of $P_0$ at each input modulator of the photonic processors and a matrix size dependent system transmission of $\alpha$ from one input modulator to one output PD, we can write the power difference between the two outputs used for balanced readout as:

$$\Delta P = P_0 \cdot \alpha \cdot \sum_{i=1}^{N} (b + x_i \cdot r) \cdot (w_{1,i} - w_{2,i})$$

*Eq. 1*

Here, $r$ is the encoding range, given by the output swing of the DAC and the response of the modulator, $b$ is the bias point transmission of the modulator and N is the matrix size. We denote the input vector as $x$, $x_i$ is within [-1,1], and the positive weights vectors contributing to the balanced readout scheme as $w_1$ and $w_2$. They are given by the absolute transmission of the analog weight representation. We can rewrite the power difference as:

$$\Delta P = P_0 \cdot \alpha \cdot \sum_{i=1}^{N} b \cdot \Delta w \cdot w_i + P_0 \cdot \alpha \cdot \sum_{i=1}^{N} x_i \cdot r \cdot \Delta w \cdot w_i$$

*Eq. 2*

Here, the memory window $\Delta w$ is the absolute difference between the high and low state of the analog weight and $w$ is the effectively encoded weight, $w_i$ is within [-1,1]. The first term is constant for a constant weight configuration and is effectively removed by the high-pass characteristic of the readout electronics:

$$\Delta P = P_0 \cdot \alpha \cdot r \cdot \Delta w \cdot \sum_{i=1}^{N} x_i \cdot w_i$$

*Eq. 3*

As designed, the difference in output power is proportional to the scalar product of the input vector and the weight vector. If we assume that both the inputs and the effective weights follow uncorrelated uniform distributions *U(-1,1)*, we can compute the expectation value and the variance of the output distribution as:

$$\langle \Delta P \rangle = P_0 \cdot \alpha \cdot r \cdot \Delta w \cdot \sum_{i=1}^{N} \langle x_i \rangle \cdot \langle w_i \rangle = 0$$

*Eq. 4*

Thus, for the variance holds:

$$\begin{aligned} \text{Var}(\Delta P) &= \langle (\Delta P - \langle \Delta P \rangle)^2 \rangle \\ &= P_0^2 \cdot \alpha^2 \cdot r^2 \cdot \Delta w^2 \cdot \sum_{i=1}^{N} \langle x_i^2 \rangle \cdot \langle w_i^2 \rangle \\ &= \frac{N}{9} \cdot P_0^2 \cdot \alpha^2 \cdot r^2 \cdot \Delta w^2 \end{aligned}$$

*Eq. 5*

Due to the central limit theorem ΔP approaches a Gaussian distribution for large values of N, e.g. already for a vector size of N = 16 as shown in Figure 5a. This distribution is quantized based on the ADC resolution, which we assume to be 8 bits. Beyond quantization, we can also define the output range, determining the values at which we clip the distribution.

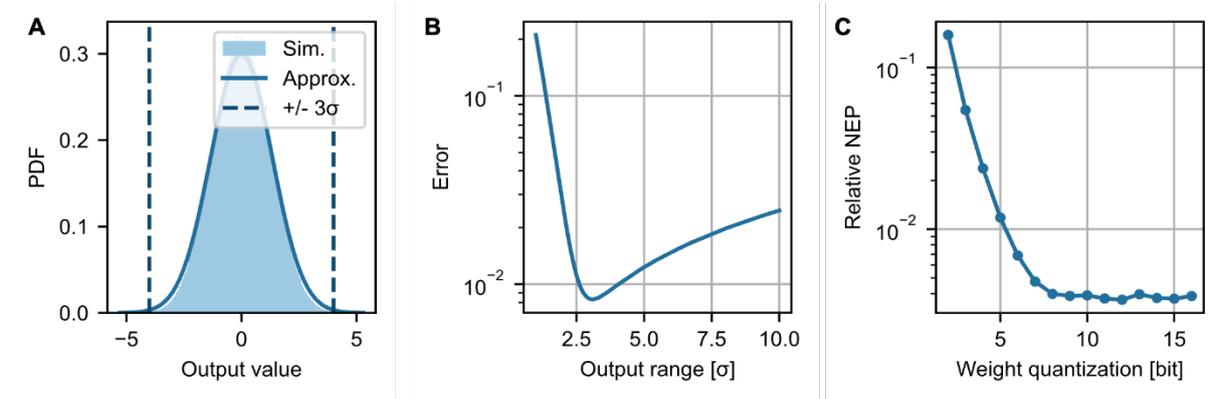

*Figure 6. Output quantization. A, As the input and weight distributions are uncorrelated, the output distribution is Gaussian already for a small matrix size of N=16. 99.7% of all values fall within 3 σ around the mean. B, Clipping the possible output range improves the quantization resolution for most of the values. For an 8bit quantization, an output range of 3.1 σ is optimal with an induced quantization error of 0.83%. C, Using the output quantization scheme, we compute the $L_2$ error for digital 64x64 matrix multiplications with 8bit input/output quantization and n-bit weight quantization. The noise equivalent optical output power of the analog system must be below 2.3% of the signal swing to achieve a 4-bit weight quantization like digital performance.*

The scalar product of *x* and *w* falls within [-N,N]. However, since the output follows a Gaussian distribution with a standard deviation of sqrt(N)/3, 99.7 % of all values lie within the [-sqrt(N),sqrt(N)]. Clipping the distribution reduces error by decreasing the step size imposed by 8-bit quantization. Figure 5b shows the error for different output ranges, using the $L_2$ norm as a measure computational accuracy:

$$L_2 = \frac{\langle |y_{\text{output}} - y_{\text{target}}| \rangle}{\langle |y_{\text{target}}| \rangle}$$

*Eq. 6*

Therefore, clipping the distribution to 3.1 σ is optimal in the given scenario. Finally, we compute the required input power $P_0$ at each modulator to achieve the target optical output signal swing:

$$P_{swing} = 3.1 \cdot \sigma = 3.1 \cdot \frac{\sqrt{N}}{3} \cdot P_0 \cdot \alpha \cdot r \cdot \Delta w$$

*Eq. 7*

And therefore:

$$P_0 = P_{swing} \cdot \frac{3}{3.1 \cdot \alpha \cdot r \cdot \Delta w \cdot \sqrt{N}}$$

*Eq. 8*

Notably, the total input power $N \times P_0$ increases with increasing vector size even for an ideal, lossless photonic processor due to the uncorrelated input distributions.

Deploying this input/output quantization scheme, we can compare the $L_2$ error of the analog system with the one of a digital system deploying the same input/output quantization but n-bit quantized weights. Modelling all analog noise sources via a noise equivalent optical output power $P_{NEP}$, the $L_2$ error of the analog system is:

$$L_2 = \frac{\langle|P_{NEP} \cdot N(0,1)|\rangle}{\langle|P_{swing}/3.1 \cdot N(0,1)|\rangle} = \frac{3.1 P_{NEP}}{P_{swing}}$$

*Eq. 9*

Note that the factor 3.1 only maps the output range/optical output signal swing back to the standard deviation of the optical output signal, since the signal to noise ratio is the crucial parameter for the computation accuracy. Figure 5c shows the equivalent weight quantization for different ratios between the $P_{NEP}$ and $P_{swing}$. In the given scenario, $P_{swing}$ must be 43.5x larger $P_{NEP}$ to achieve the computation error as a digital system with 8bit input/output quantization and 4bit weights.

Interface

The encoding mismatch between the analog photonic processor and the digital electronic system architecture requires a comparable complex interface. The interface not only consumes area and power but also introduces noise, which in turn impacts the required optical signal strength at the photonic processor's output. At the input, PAM DACs convert digital integer values into analog pulse amplitudes using architectures such as resistive ladder networks, current-steering designs, or hybrids of both. The power consumption of DACs consists of two main components: static power, for example, from continuously biasing current sources even when not actively converting; and dynamic power, such as that required to set the switches in a current-steering architecture for each input. As a result, the total power consumption of DACs does not scale linearly with increasing sampling frequency, as shown for example in the survey by P. Caragiulo et al.[42]. The energy consumed per conversion step may even slightly decrease at higher sampling rates. We approximate the energy efficiency to be (15 ± 5) fJ per conversion step at 1 GS/s and (16 ± 5) fJ per conversion step at 10 GS/s[43–46]. The physical size of the DAC typically increases with sampling rate due to the use of multiplexing techniques, which duplicate the data converter block to achieve higher throughput. At the output of the photonic processors, photodiodes couple the optical, power encoded, signals back to electronic currents. Before passing the signal to the ADCs, transimpedance amplifiers (TIAs) convert the current signals to voltages and amplify them. The performance characteristics of the TIA strongly depend on the properties of the photodiode, the ADC and the wiring between the three components. In the following we assume a TIA power consumption, which depends on both the high gain amplifier and the output buffer to drive the output (ADC) load, of (30 ± 10) mW and a size of (0.08 ± 0.02) mm². The noise equivalent input current of the TIA depends on the bandwidth, as for example the thermal noise is frequency dependent. The output signal of the photodetector must be much larger, i.e. x50, to enable high precision analog computing. Considering the TIA noise values in Table 2, output signals in the order of (20 ± 10) µA$_{pp}$ at 1 GS/s operation and (75 ± 25) µA$_{pp}$ at 10 GS/s are required. Thus, high gain amplifiers are crucial to make use of the full ADC voltage range[8,47–49]. Several different architectures exist to convert analog output voltages of the TIA back into digital values, such as flash ADCs, successive-approximation (SAR) ADCs, and voltage-controlled oscillator (VCO)-based ADCs. Despite variations in architecture and performance, these ADCs generally share the characteristic that dynamic power consumption dominates. As a result, the energy per conversion step tends to remain relatively constant at lower sampling rates and begins to increase beyond 1 GS/s, as shown in the survey by B. Murmann et al[50]. This increase is due to challenges in RF circuit design and the use of interleaving techniques, which again require duplicating data converters and increase overall power and area. For performance estimation, we assume SAR ADCs with an energy consumption of (15 ± 5) fJ per conversion step at 1 GS/s and (50 ± 15) fJ per conversion step at 10 GS/s[51–54]. Similar to DACs, the size of ADCs tends to grow with increasing sampling frequency due to the implementation of interleaved architectures.

|  | Area/mm² | Efficiency | Comments |
|---|---|---|---|
| DAC 1 GS/s[43,46] | 0.04 ± 0.01 | (15 ± 5) fJ/cs | Typical output voltage range in the order of 1 $V_{pp}$ |
| DAC 10 GS/s[44,45] | 0.05 ± 0.01 | (16 ± 5) fJ/cs | |
| TIA 1 GHz[8,48] | 0.08 ± 0.02 | (30 ± 10) mW | Typical noise equivalent input current of (0.4 ± 0.2) µA |
| TIA 10 GHz[47,49] | 0.08 ± 0.02 | (30 ± 10) mW | Typical noise equivalent input current of (1.5 ± 0.5) µA |
| ADC 1 GS/s[52,54] | 0.06 ± 0.02 | (15 ± 5) fJ/cs | Typical input voltage range in the order of 0.5 $V_{pp}$ |
| ADC 10 GS/s[51,53] | 0.18 ± 0.04 | (50 ± 15) fJ/cs | |
| Laser[55,56] | 0.1 ± 0.03 | (9 ± 3) % | Typical output power below 100 mW |
| EOM[57,58] | 0.04 ± 0.01 | (83 ± 10) % | Flat frequency response up to 10 GHz, 3dB bandwidth can exceed 100 GHz |
| PD[59,60] | 0.01 ± 0.003 | (0.9 ± 0.2) A/W | |

*Table 2. Interface components performance assumption. We use these device characteristics to simulate the scalability, computational efficiency and computational density of integrated linear photonic processors. The memory properties are architecture dependent since different computation schemes require different memory types.*

In addition to the digital-analog conversions, the system also performs electro-optic conversions. Coherent lasers, or in specific cases incoherent light sources, generate optical carrier signals. Integrated photonic lasers typically deploy indium phosphide (InP) as the gain medium for infrared operation and rely on electrical pumping. In pure InP photonic circuits, these lasers can produce up to 250 mW of output power with conversion efficiencies around 35%[55]. However, integrating InP lasers directly onto the silicon-on-insulator (SOI) platform poses challenges due to mismatches in lattice structure and thermal expansion, which eventually decreases the conversion efficiency and maximum output power[61]. For example, wafer scale flip-chip bonding of prefabricated InP lasers onto SOI enables output powers up to 40 mW with a conversion efficiency in the order of 9 % [56]. Next, the analog output pulses of the DAC are coupled to the optical carrier signal. One approach are electro absorption modulators (EAMs), which deploy fast absorption tunable waveguides, e.g. by leveraging the quantum-confined Stark effect in silicon germanium structures[62]. The second approach are electro-optic modulators (EOMs), which deploy fast phase shifters, e.g. by making use of the Pockels effect, within a Mach-Zehnder interferometer structures[63]. In the following we consider biased EOMs, their transmission depending on the input voltage V is:

$$T(V) = \sin^2\left(\frac{\pi}{4} + \frac{\pi V}{2V_\pi}\right) = \frac{1}{2} + \frac{\pi V}{2V_\pi} - \frac{\pi^3 V^3}{12 V_\pi^3} + O(V^5)$$

*Eq. 10*

$V_\pi$ is the voltage to fully open / close the EOM. In order to ensure an input encoding error below 2%, the input voltage must be within ± 0.2 $V_\pi$. Assuming an output DAC voltage of 1 $V_{pp}$, the target $V_\pi$ is around 2.5 V. Consequently, the linear encoding range is ± 0.3. Typically, there is an anti-proportional dependency between $V_\pi$ and the length of the EOM. Thus, it can be tuned by the actual circuit design. One compact way of building EOMs on SOI is integrating materials with a large $\chi^2$ non-linearity into the circuit. For example, silicon organic hybrid modulators can exhibit bandwidths beyond 40 GHz while featuring a low insertion loss below 1 dB and a modulator length below 500 µm with a $V_\pi$ in the order of 1.5 V [57,58]. At the output of the photonic processor, photodiodes couple the optical signal back to the electronic domain. Germanium is the natural choice for detection on SOI at telecom wavelengths, enabling photodiodes with 100 GHz of bandwidth and a responsivity of (0.9 ± 0.2) A/W while maintaining a compact footprint, that mainly depends on the electrode and bond pad design[59,60]. Considering the photodetector responsivity and the assumptions stated above, an optical output signal swing of (22 ± 12) µW$_{pp}$ is required for 1 GS/s operation and (83 ± 32) µW$_{pp}$ for 10 GS/s operation.

Performance Estimation

We estimate the computational density and efficiency of different photonic processor types computing NxN matrix vector multiplications. We do not consider the additional reference column in incoherent MZI meshes and photonic crossbar arrays for the power and area calculations since the impact is negligible for larger matrix sizes. When operating at a sampling frequency f with one sample per symbol, the number of operations (multiplications and additions) per second is:

$$TOPS = 2 \cdot N^2 \cdot f$$

*Eq. 11*

The total area is given by the space for the interface $A_{io}$ and the space for the photonic circuit $A_{PIC}$. The interface is identical for all processor architectures and only depends on the sampling frequency:

$$A_{io} = N \cdot (A_{DAC} + A_{EOM} + A_{Laser} + A_{PD} + A_{TIA} + A_{ADC})$$

*Eq. 12*

We can write the total power consumption as the contribution of three different uncorrelated sources

$$P = P_{el} + P_{Laser} + P_{PIC}$$

*Eq. 13*

The power consumption of the photonic integrated circuit $P_{PIC}$ is different for each architecture and depends on the matrix weight implementation. The power consumption of the electronic interface components is:

$$P_{el} = N \cdot (P_{DAC} + P_{TIA} + P_{ADC})$$

*Eq. 14*

The total required laser power is N times the laser power per port shown in Eq. 8. Also considering the conversion efficiency c of the laser source, we can write the power consumption for optical carrier generations as:

$$P_{Laser} = P_{swing} \cdot \frac{3 \cdot \sqrt{N}}{3.1 \cdot \alpha \cdot r \cdot \Delta w \cdot c}$$

*Eq. 15*

For our analysis, we assume that the EOM has a sufficiently low $V_\pi$ such that we can use the full linear encoding range of r = 0.3. We use specific measurement values for the system transmission and memory window and assume the estimates for the laser conversion efficiency and optical output signal swing discussed before. With that we can write the variance of the required laser pump power as:

$$\text{Var}(P_{Laser}) = \left(\frac{3 \cdot \sqrt{N}}{3.1 \cdot \alpha \cdot \Delta w \cdot r}\right)^2 \cdot \left(\frac{\text{Var}(c) \cdot \langle P_{swing}\rangle^2}{\langle c\rangle^4} + \frac{\text{Var}(P_{swing})}{\langle c\rangle^2}\right)$$

*Eq. 16*

**Acknowledgements**

We thank Jochen Stuhrmann, from Illustrato, for his assistance with the illustrations. The research is funded by: European Union's Horizon 2020 research and innovation programme (grant no. 101017237, PHOENICS project) and the European Union's Innovation Council Pathfinder programme (grant no. 101046878, HYBRAIN project) and European Research Council Starting Grant INFUSED.

## Data availability

The data that support the findings of this study are available from the corresponding author upon reasonable request.